\documentclass[12pt]{article}
\usepackage{a4}
\usepackage{epsf}
\usepackage{amssymb}
\usepackage{cite}
\newcommand{\be}{\begin{equation}}
\newcommand{\ee}{\end{equation}}
\newcommand{\bea}{\begin{eqnarray}}
\newcommand{\eea}{\end{eqnarray}}

\begin{document}
%%%%%%%%%%%%%%%%%%%%%%%%%%%%%%%
\def\C{{\mathbb{C}}}
\def\R{{\mathbb{R}}}
\def\s{{\mathbb{S}}}
\def\T{{\mathbb{T}}}
\def\Z{{\mathbb{Z}}}
\def\W{{\mathbb{W}}}
\def\Bbb{\mathbb}
\def\BZ{\Bbb Z} \def\BR{\Bbb R}
\def\BW{\Bbb W} 
\def\BM{\Bbb M} 
\def\e{\mbox{e}}
\def\BC{\Bbb C} \def\BP{\Bbb P}
\def\CP{\BC\BP}
%%%%%%%%%%%%%%%%%%%%%%%%%%%%
\begin{titlepage}
\title{On Dimer Models and Closed String Theories}
\author{}
\date{% authors are dated
Tapobrata Sarkar
\thanks{\noindent E--mail:~ tapo@iitk.ac.in}
\vskip0.4cm
{\sl Department of Physics, \\
Indian Institute of Technology,\\
Kanpur 208016, India}}
\maketitle
\abstract{We study some aspects of the recently discovered connection
between dimer models and D-brane gauge theories. We argue that dimer 
models are also naturally related to closed string theories on  
non compact orbifolds of $\BC^2$ and $\BC^3$, via their twisted sector 
R charges, and show that perfect matchings in dimer models 
correspond to twisted sector states in the closed string theory. 
We also use this formalism to study the combinatorics of some unstable 
orbifolds of $\BC^2$. 
}
\end{titlepage}

\section{Introduction}\label{intro}

The striking connection between dimer models of statistical mechanics 
(for reviews, see \cite{kast},\cite{kenyon}) and 
D-brane gauge theories has been the subject of much interest in the
last couple of years. The connection, first proposed in \cite{han1}
was subsequently studied in great details, and it is by now clear that
at least for the case of toric singularities, dimer models provide
the most robust computational tool for D-brane gauge theories probing
the same \cite{han2},\cite{han3}. This is important especially in the
light of  the recent extension of Maldacena's AdS/CFT correspondence 
to $AdS_5 \times X^5$ spaces, where $X^5$ is a Sasaki-Einstein manifold.
Although it seems natural that dimer 
models arise in the connection between quiver gauge theories living 
on D-brane probes, it has a more direct connection to string theory, 
which was elucidated, via mirror symmetry, in the beautiful work 
of \cite{vafa1}. 

Broadly speaking, the dimer model of a given toric singularity, whose
physical interpretation is a collection of NS5 and D5 branes, 
is obtained by constructing the dual graph of the quiver gauge theory 
living on the world volume of the D-branes probing the singularity. 
\footnote{Strictly speaking, this is called the brane tiling model,
of which each edge in a perfect matching is a dimer, but we will 
loosely refer to the dimer models and brane tiling models in the same 
spirit in what follows.}
Given such a dimer model, one can compute certain graph theoretical 
quantities, which can be shown to relate to the geometry of the singularity  
being probed by D-branes. The quiver gauge theory of a toric singularity
being a purely open string construction, it may seem that computations
from dimer models relate solely to open string quantities. However, 
the geometry of toric orbifolds has an equivalent description in 
closed string theory, and it is tempting to ask whether one can 
relate these graph theory quantities to closed strings.  
 
Indeed, it has been shown
that there is a correspondence between dimer models and the gauged
linear sigma model \cite{wittenphases} (called GLSM in the sequel). 
The dimer-GLSM correspondence was proposed in \cite{han2} and proved in 
\cite{francovegh}. Simply stated, this correspondence states that there is 
a one to one relationship between the perfect matchings in the dimer
model corresponding to the gauge theory probed by a D-brane transverse
to a given toric singularity, and the fields in the GLSM that describes
the (resolution of the) singularity. Given the fact that 
there exists purely closed string
descriptions of these singularities which also naturally relate to
the GLSM, it is probably not surprising that
there is indeed a connection between dimer models and closed strings, 
and the purpose of this paper would be to explore and specify such a 
connection, and in some sense this is complementary to the results 
of \cite{francovegh}.

A crucial aspect of the results of \cite{han1} (and the subsequent 
papers that developed these ideas more fully) is the fact that the 
combinatorics of dimer models
predict the ``multiplicities'' of the GLSM fields that arise in the
description of gauge theories living on D-branes probing toric 
singularities. These multiplicites have no obvious analogue in the
closed string picture of the resolution of these singularities. However,
as we will show in the sequel, perfect matchings of dimer models 
(explained in the next section) for toric singularities can be shown to 
specify the twisted sector R-charges of the closed string CFT states 
that describes the resolution of the singularity. These charges, 
which specify the GLSM that describes the singularity 
therefore provide a correspondence between dimers and GLSMs, via
closed strings. 

Further, we also propose a different method for the counting of perfect
matchings corresponding to a given dimer model, which is different
from the ones used in \cite{han1},\cite{han2}. Our method of counting
directly uses the R charges of the twisted sectors of closed string
theories living on orbifolds, and in a sense gives a direct interpretation
of the perfect matchings as GLSM fields.
The advantage of this construction is twofold. Apart from providing a new
perspective on dimers, it also makes the computation of perfect matchings
(and hence multiplicities of GLSM fields) simpler, and puts it in 
a broader framework. 

The paper is organised as follows. In the next section, we briefly review
the relevant results on the physics of dimer models, that will set the
notations and conventions used in the rest of the paper. In section 3,
we examine the dimer model and specify its connection to closed string
theories for the known $\BC^2$ orbifold singularities. Sections 4 deals
with singularities corresponding to orbifolds of $\BC^3$. Finally, we
end the paper in section 5, with some comments on generic 
(non-supersymmetric) orbifolds and their connections with graph theory.

\section{Dimer Models and D-brane Gauge Theories on Orbifolds}

In this section, we review various known results on dimer models
and its relationship with D-brane gauge theories on orbifold 
singularities. This will also serve to 
set the notations and conventions used in the rest of the paper.  

\subsection{D-brane Gauge Theories on Orbifolds}

To begin with, we briefly review the construction of quiver
gauge theories on D-branes probing orbifold singularities. Since 
the issue has been studied in great details over the past few years, 
we will be brief here, and provide a schematic overview of the topic. 

The toric data corresponding to a certain quotient
singularity can be described in terms of the $U(1)^N$ gauge theory of 
D-branes probing the singularity. The gauge theory is constructed by
the prescription due to Douglas and Moore \cite{dm}. 
While the pioneering work of \cite{dm},\cite{jm} was concerned with  
D-branes probing (arbitrary) ALE spaces, various results have emerged 
over the years that have generalised and extended these results to generic 
toric varieties (for a nice review on the subject and further
references, see \cite{he}). Two paradigms have emerged since then
regarding the construction of D-brane gauge theories on toric
varieties : the forward and the inverse algorithms, which have 
been studied and refined over the years and culminated in the fast 
forward and the fast inverse algorithms, using dimer models. 
Let us start with a brief description of the forward algorithm. 

The forward algorithm deals with computation of the moduli space
of a given quiver gauge theory (and a superpotential). Consider,
eg. the low energy limit of a probe D-brane gauge theory on an orbifold
singularity. The action of the discrete orbifolding group on the
coordinates and the Chan Paton factors determine the degrees of 
freedom that survive the orbifolding action, and the quiver diagram
encodes all the information about the charges of the unprojected 
fields. We will be mostly concerned with abelian orbifolds, and in 
these cases, the $U(1)^N$ charges (where $N$ denotes the number of
gauge groups modulo an overall $U(1)$ denoting the centre of mass
motion of the D-branes) is written as a matrix, $\Delta$. 

Now, the F-term (superpotential) equations of the theory, which are not 
all independent, are solved in terms of the minimal number of
independent fields. The solution is expressed in terms of a matrix
$K$, such that the original fields in the quiver (denoted by 
$X_i$) are expressed in terms of the independent fields $v_j$ 
as $X_i = \prod_j v_j^{K_{ij}}$. The $K_{ij}$ might have negative
entries, and in order to avoid possible singularities which might
arise due to this, one introduces a new set of fields $p_a$, that are
dual to the $v_i$s, by computing the dual matrix, $T$, of $K_{ij}$,
with $T$ being such that $K.T \geq 0$ for all the entries. In terms
of these fields, the original $X_i$ are solved as
$X_i =  \prod_a p_a^{K.T}$. The number of fields $p_a$ is not
determined apriori, and this leads to multiplicities in 
the dual description. Typically, the number of $p_a$s is more than the number
of $X_i$s, and for this reason, we need to introduce a certain number
of $C^*$ actions in order to eliminate redundancies. The charges of 
the $p_a$s under the new set of $C^*$ actions are determined using gauge 
invariance conditions. Further, one can determine the charges of the
$p_a$s under the original gauge group, using the matrix $\Delta$. 
These two sets of charges, when
concatenated, gives rise to a charge matrix whose kernel gives the
geometric data for the resolution of the singularity being probed by
the D-branes (note that for $N$ D-branes probing the singularity, we
get $N$ copies of the probed geometry). This procedure, first 
pioneered in \cite{dgm} gives us a generic method of obtaining the
geometric data corresponding to the gauge theory living on a D-brane
probing an orbifold singularity. \footnote{The method can also 
be applied to orbifolds whose action break space-time supersymmetry 
\cite{ts1}.}

The reverse procedure, i.e construction of the gauge theory data from
the geometric data of the singularity being probed, is what is known as
the inverse algorithm, first proposed in \cite{fhh}. The universal
method of obtaining the gauge theory data from the geometric one 
proceeds via partial resolution of abelian threefold singularities. 
The given singularity is first embedded into a generic singularity
of the form $\BC^3/\BZ_n\times \BZ_m$ (with $n$ and $m$ assumed to
be of the minimal values) and by partial resolution of the latter, 
which leads to the singularity in question, one is able to construct
the corresponding gauge theory. From the discussion of the previous
paragraph, it is obvious that there are various redundancies involved
in the process, and the resulting theory is non-unique. However, these
flow to the same universality class in the infrared, and this has
been called toric duality and has been shown to be equivalent to
Seiberg duality.  

The forward and inverse algorithms mentioned in the last two paragraphs
have been refined into the fast forward and the fast inverse algorithm
by Hanany and his collaborators via the striking
connection between gauge theories living on D-brane probes and 
certain graph theoretical models, known as dimers. The connection 
arises from the observation that the multiplicities of the GLSM fields
that we have mentioned can be determined from a graph that is in some
sense dual to the quiver diagram of the gauge theory. The main result
of the exercise is that the information about the quiver gauge theory 
living on D-branes probing a toric singularity is encoded in certain
dual graphs. This surprising connection has since been exploited to
study various issues relating to gauge theories, via dimer models. 
Before we elaborate on this, let us briefly recapitulate the essential
features of dimer models.

\subsection{Dimer Models and Orbifolds}  

In this subsection we will summarise a few basic features of 
dimer models. 
Broadly, dimer models refer to the statistical mechanics of bipartite
graphs, i.e graphs which have the property that each vertex can be
colored black or white, such that no two vertices of the same color 
are adjacent. Given such a graph, a perfect matching denotes a 
subset of edges, (called dimers), such that each vertex is the endpoint 
of precisely one edge. There can be various possible perfect matchings 
corresponding to a given bipartite graph, and the statistical mechanics 
of random perfect matchings have been the subject of much interest.  

\begin{figure}[h]
\centering
\epsfxsize=2.5in
\hspace*{0in}\vspace*{.2in}
\epsffile{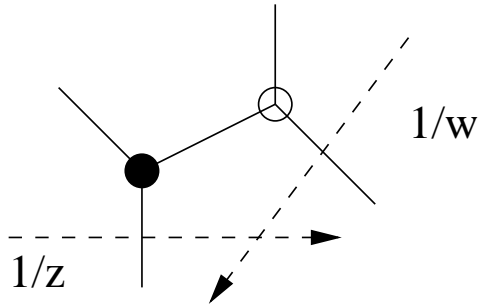}
\caption{\small The fundamental domain for the hexagonal graph} 
\label{fig0a}
\end{figure}

In \cite{han1}, a connection between the combinatorics of dimer models 
and D-brane gauge theories was proposed. Essentially, the connection arises
when one considers the ``Kasteleyn matrix'' for a given brane tiling 
obtained by dualising the (periodic) quiver diagram of the gauge theory.
\footnote{In this construction, nodes, arrows and plaquettes of the periodic 
quiver gets related to the faces, edges and nodes respectively of
the brane tiling.}  
The determinant of this matrix, called the charactaristic polynomial,
captures the multiplicities
of the GLSM fields that appear in the D-brane probe description of the
singularity. The computation of the Kasteleyn matrix has been extensively
dealt with in \cite{han1},\cite{han2},\cite{han3}. 
Essentially, for a $T^2$ embeddable graph, we can derive
the Kasteleyn matrix by constructing paths that wind around the two
cycles of the torus, and appropriately weighing the edges that are
crossed by these paths. This construction depends on the 
fundamental domain of the graph. Consider,  
e.g the hexagonal graph whose fundamental domain is shown in fig. 
(\ref{fig0a}). It can be shown \cite{han1} that the Kasteleyn matrix 
in this case is $1\times 1$ and its determinant is 
\be
P\left(z,w\right) = 1 - z - w
\ee
%%%%%%%%%%%%%%%%%%%%%%%%%%%%
%%%%%%%%%%%%%%%%%%%%%%%%%%%%
The hexagonal graph in fact corresponds to $\BC^3$ and an arbitrary orbifold
of $\BC^3$ is obtained by taking copies of the fundamental domain. The 
procedure is standard, and it can be shown that the determinant of the
Kasteleyn matrix correctly reproduces the multiplicities of the GLSM fields
for D-branes probing these orbifolds. This has led to the 
dimer-GLSM correspondence, which states that every perfect matching in the
dual graph of the quiver gauge theory of a toric singularity is in one to
one correspondence with the fields in the GLSM construction of the toric
moduli space of the singularity. The conjecture was put forward in \cite{han2}
and subsequently proved in \cite{francovegh}. 

Having reviewed the essential features of dimer models, we now briefly 
discuss the issue of closed string theories on orbifold singularities.   
 
\subsection{Closed Strings and Orbifolds}

In this subsection, we will consider closed strings on 
two fold orbifolds of the form $\BC^2/\BZ_n$ and three folds 
of the form $\BC^3/\BZ_n$ and
$\BC^3/\BZ_n\times\BZ_m$. Let us start with the $\BC^2/\BZ_n$ example. 
The orbifolding action, in this case, is given by
\begin{equation}
\left(Z_1, Z_2\right) \to \left(\omega Z_1, \omega^p Z_2\right)
\end{equation}
where $p$ is an integer coprime to $n$ with $n > p > 0$, and $\omega$ is
the nth root of unity. Following standard conventions, we will denote this 
orbifold as $\BC^2/\BZ_{n(p)}$. For $p = n-1$, the orbifolding action preserves
space-time supersymmetry, while for generic values of $p \neq n-1$, the theory 
contains tachyons localised at the orbifold fixed point, and space-time
supersymmetry is broken. 

The orbifold twisted chiral ring is made out of the twist operators 
$\mathcal{T}_j,~j=1,\cdots n-1$, and are given by
\begin{equation}
\mathcal{T}_j = \mathcal{T}^x_{\frac{j}{n}}\mathcal{T}^y_{\frac{jp}{n}}
\label{rcharges}
\end{equation}
where $x$ and $y$ label the two $\BC^2$ directions.
These operators correspond to the $(c,c)$ ring of the orbifold. There is 
another set of operators which is BPS under a different combination of 
supersymmetries. These are projected out in Type II theories for 
supersymmetric orbifolds. 
\footnote{For space-time non-supersymmetric orbifolds, 
GSO projection in Type II theories might result in some of the $(c,c)$ ring 
operators being projected out. For the purpose of this paper, we will 
restrict our attention to Type 0 theories for the case $p \neq n-1$.}
The twist operators of eq. (\ref{rcharges}) carry R charges 
$\left(\frac{j}{n},\frac{jp}{n}\right)$ and the inclusion of all the
twisted sectors of the theory constitutes a canonical resolution of the
space-time supersymmetric orbifold $\BC^2/\BZ_n$. For $p \neq n-1$, some
of the twisted sectors might become irrelevant, and one needs to consider
a blowup of the singularity using only relevant (or marginal) operators. 

Once the R-charges of the twisted sectors is specified, it is easy to 
write down the toric data for the resolution of the orbifold. This is
obtained by adding fractional points corresponding to the R-charges of
the twisted sectors of the orbifold participating in the resolution in
a unit 2-D lattice generated by vectors ${\vec e_1} = \left(1,0\right)$ 
and ${\vec e_2} = \left(0,1\right)$, and
then restoring integrality in the lattice \cite{reid},
\cite{ag}. This can be done for supersymmetric as well as non-supersymmetric
orbifolds \cite{ts2}. Let us illustrate this with an example. Consider,
eg. the non-supersymmetric orbifold $\BC^2/\BZ_{5(2)}$. The relevant
deformations are those with R-charges
\be
\left(\frac{1}{5},\frac{2}{5}\right),~~\left(\frac{3}{5},\frac{1}{5}\right)  
\ee
By adding these fractional points in a two dimensional lattice with 
generators ${\vec e_1}$ and ${\vec e_2}$,
we see that the toric data for the resolution of this singularity
is given (after restoring integrality in the lattice) by
\be
T = \pmatrix{1&0&-1&-2\cr 0&1&3&5}
\label{toric1}
\ee
The kernel of the toric data gives the GLSM charges for the resolution of
the singularity, and in this case, a choice of the charge matrix (equivalent
to the kernel of $T$ in eq. (\ref{toric1}) is 
\be
Q = \pmatrix{1&2&-5&0\cr 3&1&0&-5}
\ee
 
With this discussion, we are now in a position to connect the various 
issues addressed in this section. Clearly, gauge theories 
living on D-branes probing orbifold singularities can be addressed in a 
variety of ways, all of which relate to the Witten's GLSM. The closed
string picture uses the $(2,2)$ SCFT of the world sheet to relate the
R-charges of the latter with the toric data of the resolution of the
singularity. By resolving points in the toric diagram (i.e removing
some of its vertices), we can reach various partial resolutions of 
orbifolds. This is equivalent to giving vevs to some fields in the GLSM
description, and the procedure gives us various phases of 
D-brane gauge theories on the resolutions. From the world sheet perspective,
this is equivalent to removing some of the twisted sector charges from
the resolution. 
An important difference between the two descriptions is
that the world sheet picture for orbifolds does not capture the
multiplicities of the GLSM fields obtained in the open string picture. 

It is therefore interesting to ask if one can relate the dimer model 
description of D-brane gauge theories on orbifold singularities 
directly to closed string theories. Since there is a natural correspondence 
between closed string theories and the GLSM, and there also exists the
dimer-GLSM correspondence, a relationship between the dimer models and
closed string theories will give us a complementary approach to the
dimer-GLSM correspondence. It is this question that we will address in
the rest of the paper, for the case of orbifolds of 
$\BC^2$ and $\BC^3$. 

\section{Dimers and Closed Strings : The $\BC^2/\BZ_n$ case}

In this section, we will explore the possible connections between dimers
and Type II closed string theories on orbifolds of $\BC^2$. We will first
do a graphical analysis of the problem, by drawing the dimer models and
constructing the perfect matchings. In the next subsection, we will 
give a mathematical formulation of the partition function of the 
perfect matchings. 

\subsection{Graphical Analysis}
\begin{figure}[h]
\centering
\epsfxsize=3.5in
\hspace*{0in}\vspace*{.2in}
\epsffile{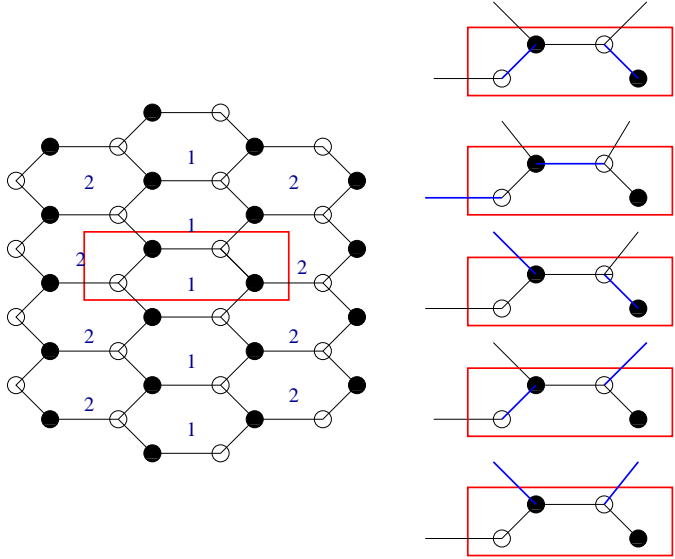}
\caption{\small The brane tiling model for the orbifold $\BC^2/\BZ_2$.
We have also shown the perfect matchings on the right.}
\label{fig1}
\end{figure}

Let us start
with the simplest example of the orbifold $\BC^2/\BZ_2$, whose brane 
tiling and perfect matchings we record for reference in fig 
(\ref{fig1}). \footnote{Throughout the paper, we will denote the 
perfect matchings with blue lines in the figures.} This orbifold has
one twisted sector, with the twisted sector R charge being given 
by $\left(\frac{1}{2},\frac{1}{2}\right)$, and this, along with the
unit vectors in two dimensions, constitute the full resolution of 
the orbifold (there is one $\BP^1$ that is blown up in the process).
In summary, the toric data for this orbifold is given by
{\small
\begin{eqnarray}
{\cal T}=
\pmatrix{
1&0&-1\cr
0&1&2\cr
}\label{data1a}\end{eqnarray}}
%%%%%%%%%%%%%%%%%%%%%%%%%%%%
%%%%%%%%%%%%%%%%%%%%%%%%%%%%
The probe D-brane gauge theory can be easily calculated, and it gives the
final toric data as in eq. (\ref{data1a}) above, with a multiplicity
of $2$ being associated to the field $\pmatrix{0\cr 1\cr}$. 

\begin{figure}[h]
\centering
\epsfxsize=5.5in
\hspace*{0in}\vspace*{.2in}
\epsffile{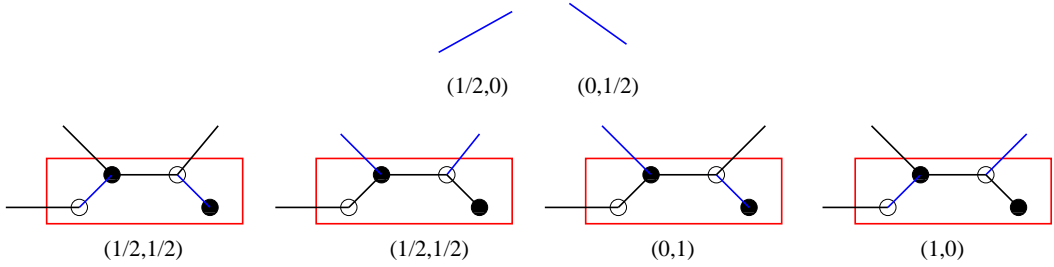}
\caption{\small The perfect matchings for the orbifold $\BC^2/\BZ_2$.
The blue lines on the top indicate the edge weights associated to the
two different edges. We have also labeled each perfect matching by
the corresponding twisted sector R charge. The redundant matching 
corresponding to the adjoint field is not shown.}
\label{fig2}
\end{figure}

This counting is reproduced in fig. (\ref{fig2}). Note that there
are two distinct edges of the graph (the third edge
that describes the redundant third direction). We weigh each edge by
the inverse of the rank of the orbifolding group, and associate a weight 
$\left(\frac{1}{2},0\right)$ and $\left(0,\frac{1}{2}\right)$ to the
two distinct edges as shown in figure (\ref{fig2}). 
\footnote{We ignore the third edge
since it gives an adjoint field that is due to the fact that the orbifold
is actually $\BC^2/\BZ_2 \times \BC$ rather that $\BC^2/\BZ_2$.} 
This counting of the perfect matchings is different from the counting
of corresponding quantities in terms of the height function 
\cite{han1}. Our counting proceeds via the edge weight, 
and directly gives us the closed string twisted sector R charges.
Hence, given the quiver gauge theory, and the brane tiling, 
we can cast the problem of counting perfect matchings to the closed
string language and can directly read off the twisted sector charges,
and their multiplicities.  
%%%%%%%%%%%%%%%%%%%%%%%%%%%%
%%%%%%%%%%%%%%%%%%%%%%%%%%%%

\begin{figure}[h]
\centering
\epsfxsize=5.5in
\hspace*{0in}\vspace*{.2in}
\epsffile{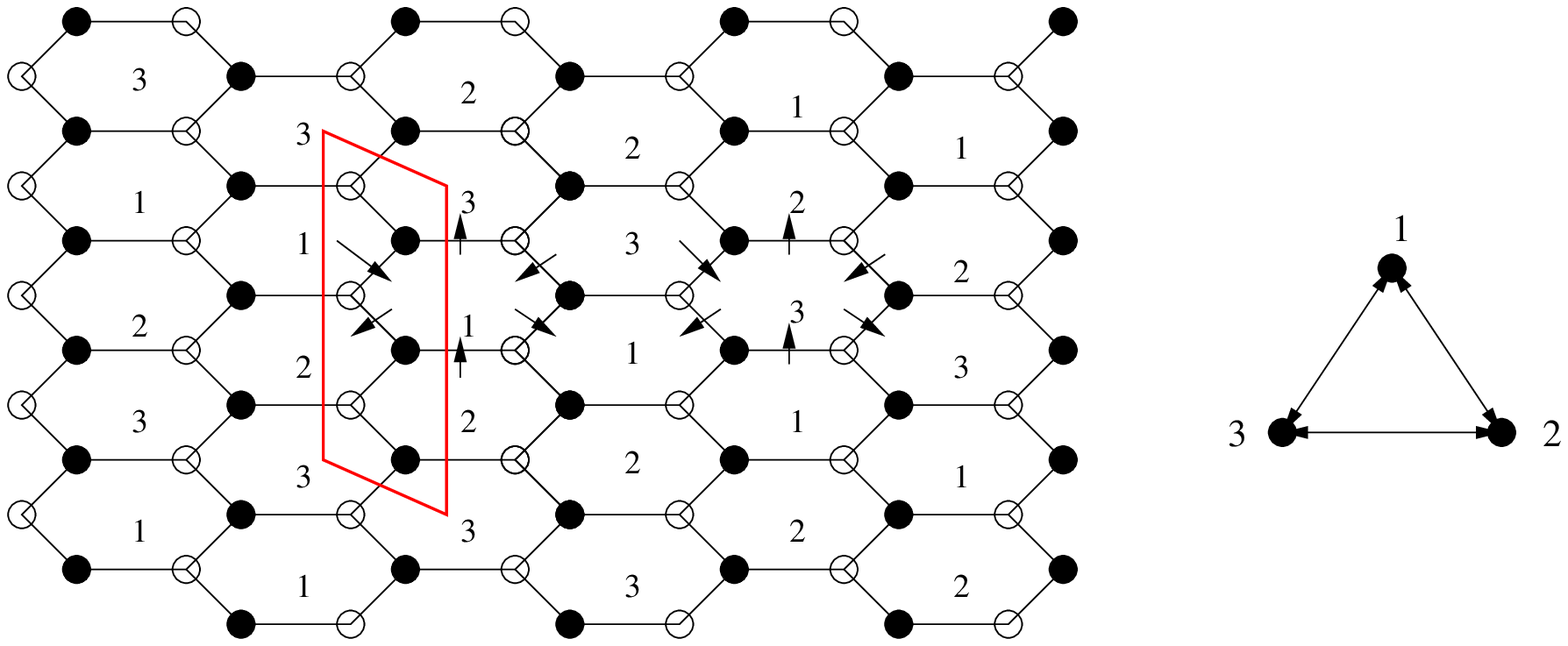}
\caption{\small The brane tiling model for the orbifold $\BC^2/\BZ_3$
and its quiver diagram without a redundant adjoint field. The fundamental
cell in the graph has been marked in red.}
\label{fig3}
\end{figure}

\begin{figure}[h]
\centering
\epsfxsize=3.5in
\hspace*{0in}\vspace*{0.2in}
\epsffile{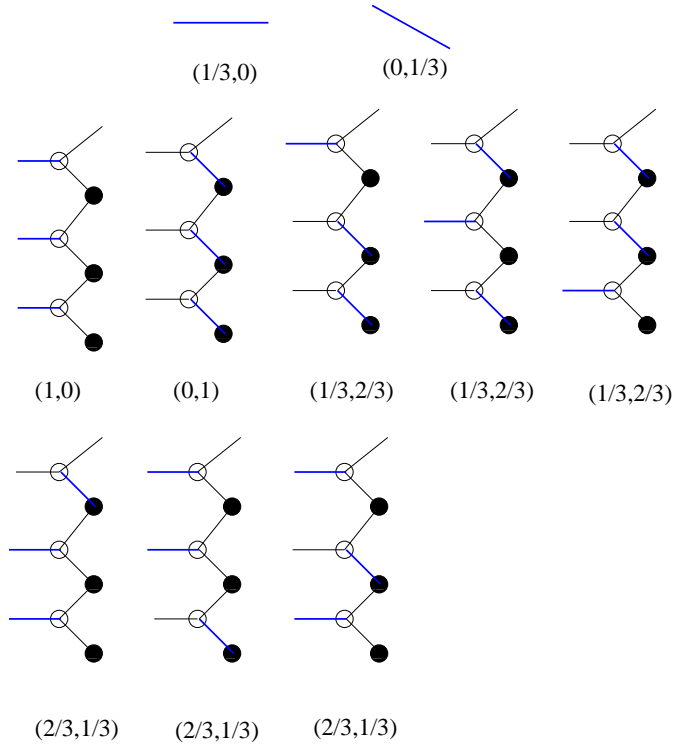}
\caption{\small The perfect matchings for the orbifold $\BC^2/\BZ_3$. We
have not included a final matching involving the redundant edge.}
\label{fig4}
\end{figure}

Let us now turn to the orbifold $\BC^2/\BZ_3$. In fig. (\ref{fig3})
we have shown the brane tiling model for this orbifold and 
its quiver diagram.  
%%%%%%%%%%%%%%%%%%%%%%%%%%%%
%%%%%%%%%%%%%%%%%%%%%%%%%%%%
Figure (\ref{fig4}) shows the perfect matchings for the orbifold
$\BC^2/\BZ_3$ in the fundamental domain of the tiling.  
%%%%%%%%%%%%%%%%%%%%%%%%%%%%
%%%%%%%%%%%%%%%%%%%%%%%%%%%%
In this case, as before, we have weighted the edges with the 
(inverse of the) rank of
the orbifolding group (except the redundant edge corresponding to the
adjoint field) and we have also labeled the perfect matchings in terms
of the closed string R charge. We see that the multiplicities of 
the $\BC^2/\BZ_3$ is reproduced in this way of counting, in which
we have not made use of the Kasteleyn matrix. 

\begin{figure}[h]
\centering
\epsfxsize=2.5in
\hspace*{0in}\vspace*{-0.2in}
\epsffile{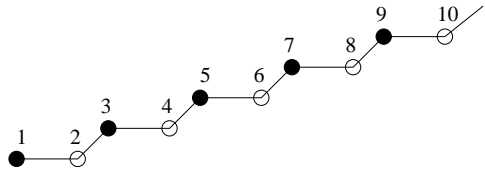}
\caption{\small The fundamental unit for $\BC^2/\BZ_5$. This contains 
ten nodes and ten edges.}
\label{fig5}
\end{figure}

\begin{figure}[h]
\centering
\epsfxsize=3.5in
\hspace*{0in}\vspace*{.2in}
\epsffile{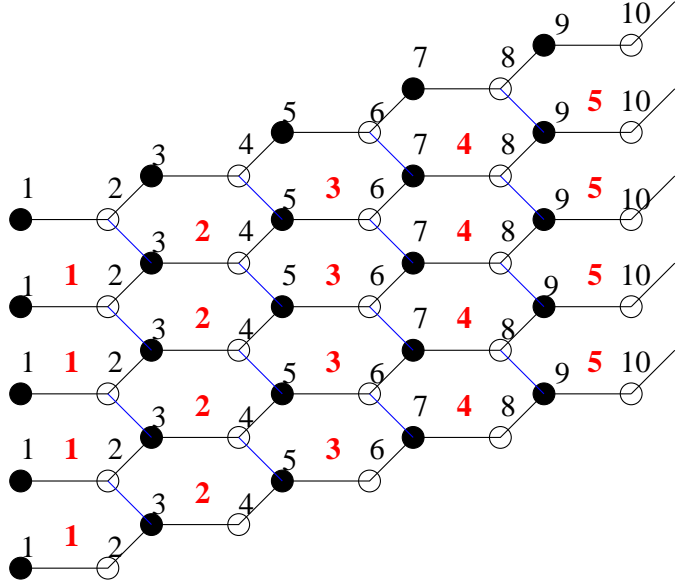}
\caption{\small Making the fundamental unit for $\BC^2/\BZ_5$ periodic,
the dimer model for this orbifold emerges.}
\label{fig6}
\end{figure}

Comparing fig. (\ref{fig2}) and 
fig. (\ref{fig4}), we notice a pattern. For any orbifold of the form
$\BC^2/\BZ_n$, with the rank of the orbifolding group being $n$, 
the fundamental domain consists of $2n$ points, with $3n$ edges, $n$
of each type, with one type being redundant, corresponding to 
the unorbifolded $\BC$ direction. This means that given an orbifold
$\BC^2/\BZ_n$, we can engineer its dimer model as follows. We take $n$ 
edges of two types, the first type having an edge weight $\left(
\frac{1}{n},0\right)$ and the other having an edge weight 
$\left(0,\frac{1}{n}\right)$, with each edge joining two different types 
of points, which we decide to color white or black. \footnote{There
exists an ambiguity here. The two edges represent the two complex
directions that are orbifoldized. Hence, we could exchange the two 
weights without affecting the results. We will have more to say about
this towards the end of this section.} We then make a 
periodic array using this fundamental unit, and this naturally gives 
the dimer model for the orbifold $\BC^2/\BZ_n$. This procedure is 
illustrated for the orbifold $\BC^2/\BZ_5$ in figs. (\ref{fig5}) and
(\ref{fig6}). In fig. (\ref{fig5}), we have illustrated the fundamental
unit for the orbifold. 
%%%%%%%%%%%%%%%%%%%%%%%%%%%%
%%%%%%%%%%%%%%%%%%%%%%%%%%%%
Fig. (\ref{fig6}) shows the emergence of the
dimer model for this orbifold on making the fundamental unit periodic.
%%%%%%%%%%%%%%%%%%%%%%%%%%%%
%%%%%%%%%%%%%%%%%%%%%%%%%%%%
Note that this gives us a direct way of relating dimer models with closed
string theories and hence to the GLSM. Let us see if we can substantiate 
this. Consider, eg. the orbifold $\BC^2/\BZ_3$. This theory has two 
marginal twisted sectors, with R charges given by
\be
\left(\frac{1}{3}, \frac{2}{3}\right), \left(\frac{2}{3}, \frac{1}{3}\right)
\ee
The toric data for this orbifold is obtained by considering the two
dimensional lattice, generated by ${\vec e_1} = \left(1,0\right)$ and
${\vec e_2} = \left(0,1\right)$ with these fractional points. Now, 
restoring integrality in the lattice gives us the toric data for the
orbifold,   
{\small
\begin{eqnarray}
{\cal T}=
\pmatrix{
1&0&-1&-2\cr
0&1&2&3\cr
}\label{data2}\end{eqnarray}}
The charge matrix for the GLSM for this orbifold is given in a 
particular basis by
{\small
\begin{eqnarray}
{\cal T}=
\pmatrix{
1&2&-3&0\cr
2&1&0&-3\cr
}\label{data2a}\end{eqnarray}}
Hence, the twisted sector R charges, which determine the dimer model for
this orbifold is directly related to the GLSM charge matrix. Note also
that as mentioned earlier, each perfect matching of the dimer model can
be taken to represent a twisted sector in the corresponding closed 
string theory, i.e a field in the GLSM. Of course, the closed 
string theory of an orbifold does not see the multiplicities associated
with the open string picture. But here we see that in terms of graph 
theory, the multiplicities actually correspond to various ways of 
constructing the twisted sector.  

Before closing this subsection, let us point out that for 
supersymmetric $\BC^2/\BZ_n$ orbifolds, our construction
naturally gives the multiplicities of the GLSM fields as coefficients
in the binomial expansion of $(x+y)^n$ and the total multiplicity
as $2^n$, consistent with \cite{han4}. 

\subsection{The Partition Function of Perfect Matchings}

We now give a mathematical description of the graphical analysis that
we have carried out in the last subsection. Let us emphasize that we 
will not be using the height function \cite{kenyon} in writing down 
this partition function, but will simply use the properties of the 
closed string twisted sectors. 

It is easy to write down 
the partition function $P$ of perfect matchings for the supersymmetric 
orbifolds $\BC^2/\BZ_n\times\BC$. We will simply write the result here.  
\begin{equation}
P = \prod_{i=0}^{n-1}\left(1 + \omega^ix^{\frac{1}{n}}
+\omega^iy^{\frac{1}{n}}\right)
\label{partition1}
\end{equation}
where the $1$ appearing in the formula represents the redundant direction
which is not taken into account while evaluating the perfect matchings, 
$\omega = e^{\frac{2\pi i}{n}}$ and we will interpret the
product as running over the $(n-1)$ twisted sectors of the orbifold
as well as the untwisted sector.
The product can be expanded, and gives the multiplicities
of the GLSM fields in accordance with \cite{han4}. In addition, it
indicates the twisted sector R-charges of the blowup modes. Eg. for
the case $n=5$, we obtain
\begin{equation}
P = 1 + x + y + 5x^{\frac{1}{5}}y^{\frac{4}{5}} 
+ 10x^{\frac{2}{5}}y^{\frac{3}{5}} + 10x^{\frac{3}{5}}y^{\frac{2}{5}} 
+ 5x^{\frac{4}{5}}y^{\frac{1}{5}}
\label{partition2}
\end{equation}
we note that the powers of $x$ and $y$ reproduce the R-charges of
the twisted sector fields. 
The additional unit term corresponds to the redundant $\BC$ direction.  
From a closed string perspective, these R charges constitute the
Newton polygon in the resolution of the singularity $\BC^2/\BZ_5$.
Note the difference in our formula of eq. (\ref{partition1}) and 
the counting of \cite{han1},\cite{han2} which uses the height function
\cite{kenyon} of perfect matchings for the counting. In this case, 
the variables $x$ and $y$ have a physical interpretation as representing the 
two complex directions of $\BC^2$. By analogy, for the case of the
$\BC^3$ orbifolds, we need three variables. We will see in the next
section how this can be nicely interpreted as a supersymmetric
completion of non-supersymmetric $\BC^2$ orbifolds. \footnote{By
supersymmetric completion, we mean that given a non-supersymmetric
$\BC^2$ orbifold of the form $\BC^2/BZ_{n(p)}$, we add an extra
direction so that the orbifolding action is now
$\left(Z_1,Z_2,Z_3\right) \to \left(\omega Z_1, \omega^p Z_2,
\omega^{n-p-1}Z_3\right)$. This is a supersymmetric $\BC^3/\BZ_n$
orbifold.} 

According to our discussion in the previous
section, for the supersymmetric $\BC^2/\BZ_5$ orbifold, the fractional
points that need to be added to the $SL(2,Z)$ lattice generated by
${\vec e_1} = (1,0)$ and ${\vec e_2} = (0,1)$ are the set of points
\be
\left(\frac{1}{5},\frac{4}{5}\right),~~\left(\frac{2}{5},\frac{3}{5}\right) 
\left(\frac{3}{5},\frac{2}{5}\right),~~\left(\frac{4}{5},\frac{1}{5}\right) 
\ee
In order to restore integrality in this lattice, we pick the points
$(\frac{1}{5},\frac{4}{5})$ and $(0,1)$ which we now call 
${\vec e_1}$ and ${\vec e_2}$ respectively. In terms of these, the precise
mapping between the twisted sector charges and the toric data is given
by 
\begin{eqnarray}
&~&\left(\frac{1}{5},\frac{4}{5}\right) \to \left(1,0\right),~~
\left(\frac{2}{5},\frac{3}{5}\right) \to \left(2,-1\right),~~
\nonumber\\
&~&\left(\frac{3}{5},\frac{2}{5}\right) \to \left(3,-2\right),~~
\left(\frac{4}{5},\frac{1}{5}\right) \to \left(4,-3\right),~~
\end{eqnarray}
With the point ${\vec e_1}$ being given by $\left(5,-4\right)$ in 
the new integral lattice. With this mapping, we can identify 
the points of the toric diagram with the corresponding twisted 
sectors, and once this is done, eq. (\ref{partition2}) reproduces
the multiplicities in the probe D-brane picture of the resolution
of $\BC^2/\BZ_5$.\footnote{ 
Clearly, there is a degeneracy in the choice of basis vectors (all
of which give the same toric data). We will always follow the convention
that lattice integrality is restored by expressing the vectors in terms
of the first twisted sector and ${\vec e_2}$. However, it can be easily
checked that any other choice of basis vectors will not affect our result.} 

As another example, consider the supersymmetric orbifold $\BC^2/\BZ_7$.
The toric data for this orbifolds is given by
\begin{equation}
\mathcal{T} = \pmatrix{1&0&-1&-2&-3&-4&-5&-6\cr0&1&2&3&4&5&6&7}
\end{equation}
For this orbifold, we use eq. (\ref{partition1}), with $n=7$ to obtain 
\begin{equation}
P = 1 + x + y + 7x^{\frac{6}{7}}y^{\frac{1}{7}} 
+ 21x^{\frac{5}{7}}y^{\frac{2}{7}} + 35x^{\frac{4}{7}}y^{\frac{3}{7}} 
+ 35x^{\frac{3}{7}}y^{\frac{4}{7}} + 21x^{\frac{2}{7}}y^{\frac{5}{7}}
+ 7x^{\frac{1}{7}}y^{\frac{6}{7}}
\label{partition3}
\end{equation}
This formula exactly reproduces the twisted sector R charges and their
multiplicities in the toric diagram of $\BC^2/\BZ_7$. 

The advantage of the formula in eq. (\ref{partition1}) is that this can
be used to compute the multiplicities of the non-supersymmetric 
$\BC^2$ orbifolds. \footnote{We comment on the analogues of the brane 
tiling models for these non-supersymmetric orbifolds towards the end of
the paper. Right now, we simply use eq. (\ref{partition1}) as a tool
to evaluate the multiplicities of these. The computation, however, is clear 
from our interpretation of the perfect matchings as twisted sector fields.}
As we have mentioned earlier, the gauge theory data
for space-time non-supersymmetric orbifolds can be calculated in much
the same way as the supersymmetric ones, following the methods of
\cite{dgm} (for non-cyclic orbifolds, see also \cite{ts3})
From our analysis of the supersymmetric orbifolds above, the procedure
is clear for non-supersymmetric orbifolds. We see that
the partition function of perfect matchings for the non-supersymmetric
$\BC^2/\BZ_{n(p)}$ orbifold is
\begin{equation}
P = \prod_{i=0}^{n-1} \left(1 + \omega^i x^{\frac{1}{n}} +
\omega^{ki} y^{\frac{1}{n}}\right)
\label{partition4}
\end{equation}
where $k < n$ is the smallest integer such that $kp = -1 ({\rm{mod}}~n)$.
Let us apply this to the orbifold $\BC^2/\BZ_{5(3)}$. The resolution of
this singularity corresponds to turning on two $\BP^1$s, with self
intersection numbers $2$ and $3$, as can be seen from the continued 
fraction $\frac{5}{3} = 2 - \frac{1}{3}$. The open string
picture of this non-supersymmetric orbifold can be obtained by following
the methods outlined in subsection (2.1). The calculations are lengthy 
to be produced here, and we simply provide the final result for the 
toric data 
\begin{equation}
T = \pmatrix{1&0&-1&-3\cr 0&1&2&5\cr {\bf 1}&{\bf 5}&{\bf 5}&{\bf 
1}\cr}
\label{toricc2z5}
\end{equation}
where, in the last row, we have also provided the multiplicities of the
GLSM fields.

The closed string picture of this orbifold consists of the two 
relevant (tachyonic) twisted sectors with R charges 
$\left(\frac{1}{5},\frac{3}{5}\right), \left(\frac{2}{5},\frac{1}{5}\right)$.
The toric data of this orbifold consists of these fractional points
in addition with the basis vectors ${\vec e_1}$ and ${\vec e_2}$. In
order to restore integrality in the lattice, we choose a new basis, and
following our convention, we call the point $\left(\frac{1}{5},
\frac{3}{5}\right)$ and $\left(0,1\right)$ as our new basis vectors
${\vec e_1}$ and ${\vec e_2}$. A simple calculation shows
that we then reproduce the data in eq. (\ref{toricc2z5}). 
The partition function of perfect matchings
for this orbifold is given by eq. (\ref{partition4}) with $n=5, k = 3$. 
Putting these values in eq. (\ref{partition4}), we obtain
\begin{equation}
P = 1 + x + y - 5x^{\frac{1}{5}}y^{\frac{3}{5}} + 
 5x^{\frac{2}{5}}y^{\frac{1}{5}} 
\end{equation}
The above formula (modulo minus signs) precisely reproduces the toric data in 
eq. (\ref{toricc2z5}) along with the relevant twisted sector charges.

As a more complicated example, let us consider the non-supersymmetric
orbifold $\BC^2/\BZ_{7(3)}$. The resolution of this orbifold corresponds
to the blowup of three $\BP^1$s with self intersection numbers 
$(3,2,2)$. In the open string language, the toric data can be calculated
using methods outlined in subsection (2.1), and we simply present the
final result
\be
T = \pmatrix{1&0&-1&-2&-3\cr 0&1&3&5&7\cr {\bf 1}&{\bf 7}&{\bf 14}
&{\bf 7}&{\bf 1}}
\label{c2z7}
\ee
where in the last line we have given the multiplicities of the GLSM
fields that appear in the resolution. This can be obtained from
our formula in eq. (\ref{toricc2z5}) with $n=7, k = 2$ and the
partition function is
\begin{equation}
P = 1 + x + y - 7x^{\frac{1}{7}}y^{\frac{3}{7}} +
14x^{\frac{3}{7}}y^{\frac{2}{7}} - 7x^{\frac{5}{7}}y^{\frac{1}{7}} 
\end{equation}
this is seen to reproduce the data in eq. (\ref{c2z7}). 

It is not difficult to interpret these results in terms of the perfect
matching that we have introduced earlier. The non-supersymmetric orbifolds
simply correspond to assigning different weights to the edges of the
fundamental cell described earlier. Consider, eg. the orbifold 
$\BC^2/\BZ_3$ of fig. (\ref{fig4}). If, instead of weights 
$\left(\frac{1}{3},0\right), \left(0,\frac{1}{3}\right)$, we assign
the weights $\left(\frac{1}{3},0\right), \left(0,\frac{2}{3}\right)$
to the edges of fig. (\ref{fig4}), we see that three of the perfect 
matchings now describe twisted sectors that are irrelevant (i.e
the sum of the R-charges exceed unity) and hence need to be projected
out of the description of the resolution of the singularity. Only
the sector with R-charge $\left(\frac{1}{3},\frac{1}{3}\right)$ 
is relevant, and this, along with the generators ${\vec e_1}$ and
${\vec e_2}$ of the two dimensional lattice describe the Newton polygon
for the resolution of the singularity $\BC^2/\BZ_{3(1)}$. 
\footnote{In some cases, we find that there it is necessary to 
interchange the roles of $x$ and $y$ in order to obtain the correct
toric data. This is attributed to the ambiguity in assigning weights
referred to earlier in this section.} 

This completes our discussion on generic orbifolds of $\BC^2$. In the
next section, we will focus on orbifolds of $\BC^3$. 

\section{Dimers and Closed Strings : The $\BC^3$ case}

In this section, we will deal with supersymmetric orbifolds of 
$\BC^3$. We will discuss cyclic as well as non-cyclic orbifolds of
$\BC^3$. We start with the cyclic orbifolds of the form 
$\BC^3/\BZ_n$.

\subsection{Cyclic orbifolds of $\BC^3$}
 
\begin{figure}[h]
\centering
\epsfxsize=5.5in
\hspace*{0in}\vspace*{.2in}
\epsffile{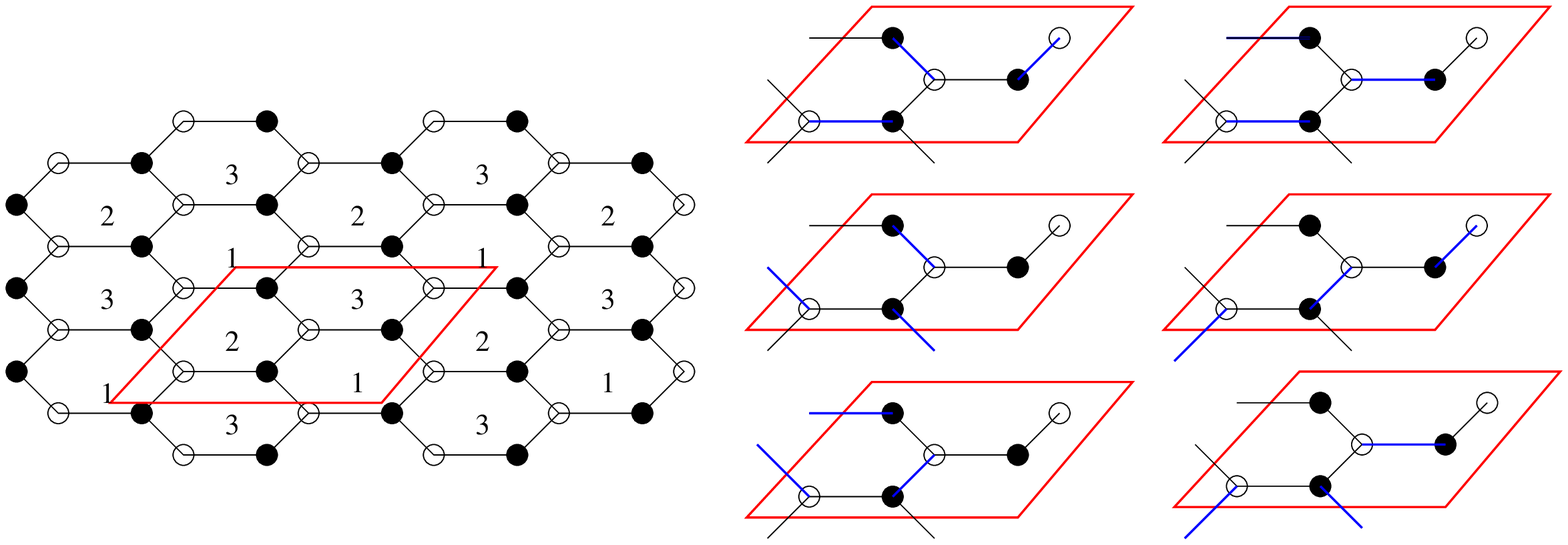}
\caption{\small The dimer model and perfect matchings for the 
supersymmetric orbifold $\BC^3/\BZ_3$. The fundamental cell is shown 
in red.}
\label{fig7}
\end{figure}

\begin{figure}
\centering
\epsfxsize=3.2in
\hspace*{0in}\vspace*{0.2in}
\epsffile{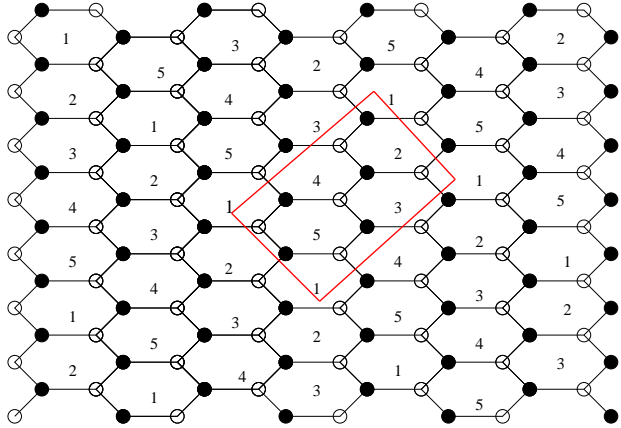}
\caption{\small The dimer model for the supersymmetric 
orbifold $\BC^3/\BZ_5$. The fundamental cell is shown in red.}
\label{fig9}
\end{figure}

\begin{figure}
\centering
\epsfxsize=6.5in
\hspace*{0in}\vspace*{.2in}
\epsffile{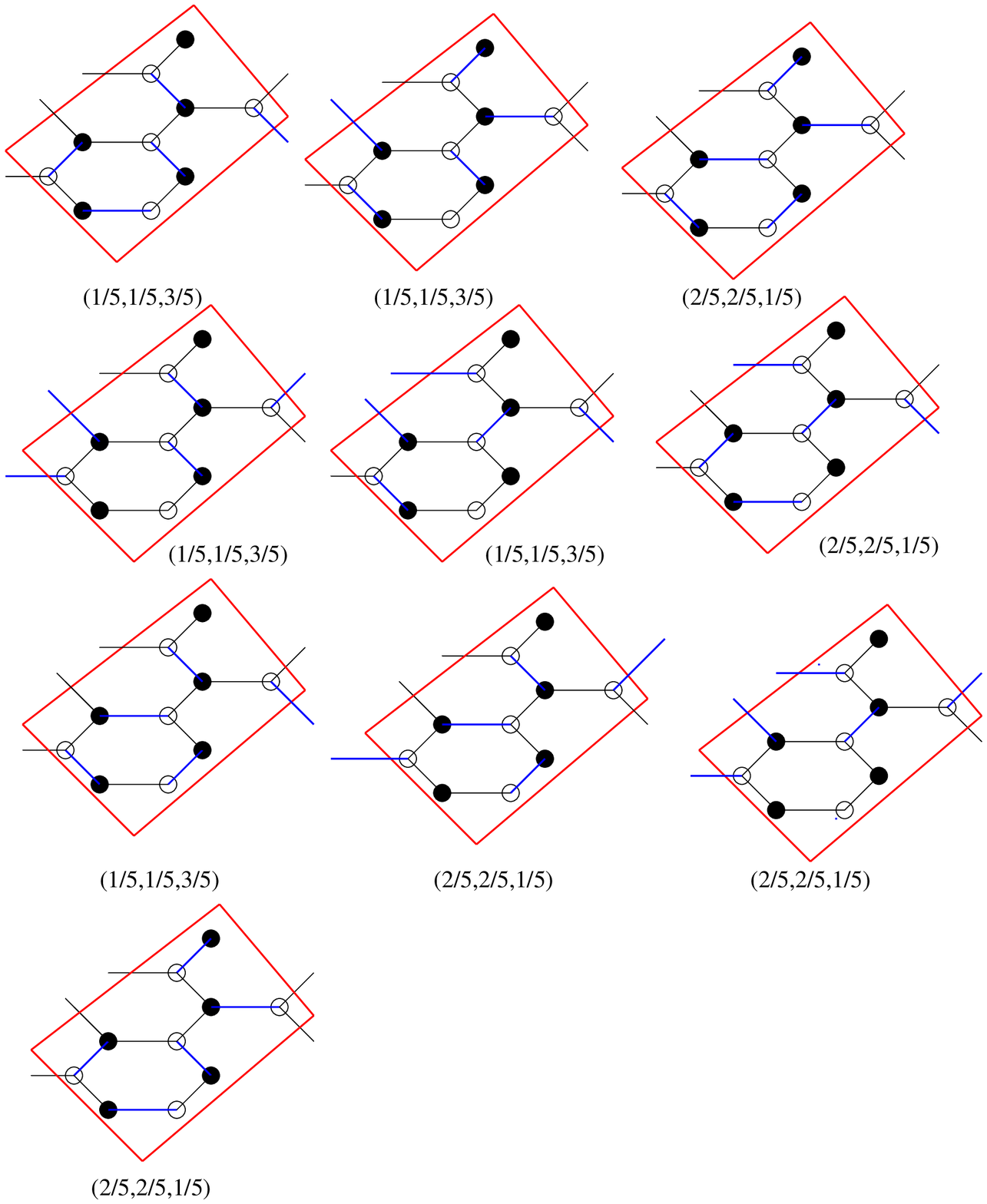}
\caption{\small The perfect matchings for the supersymmetric 
orbifold $\BC^3/\BZ_5$. The weights for the edges are 
$\left(\frac{1}{5},0,0\right), \left(0,\frac{1}{5},0\right), 
\left(0,0,\frac{1}{5}\right)$. We have not shown the unit vectors
in this construction. The red box indicates the fundamental cell
which we have retained for convenience.}
\label{fig10}
\end{figure}

Let us start with the 
supersymmetric orbifold $\BC^3/\BZ_3$. The dimer  
model and the perfect matchings of this orbifold is shown in 
fig. (\ref{fig7}). We see a familiar pattern from our analysis of
the $\BC^2$ orbifolds. Clearly, counting the number of edges with
weights $\left(\frac{1}{3},0,0\right), \left(0,\frac{1}{3},0\right)$
and $\left(0,0,\frac{1}{3}\right)$ (that can be arbitrarily assigned), 
we find that there are three perfect matchings 
corresponding to the twisted sector given by the R-charges
$\left(\frac{1}{3},\frac{1}{3},\frac{1}{3}\right)$ and the others
represent the generators of the unit three dimensional lattice.  
%%%%%%%%%%%%%%%%%%%%%%%%%%%%
%%%%%%%%%%%%%%%%%%%%%%%%%%%%
A similar exercise can be carried out for the case of the 
orbifold $\BC^3/\BZ_5$. The toric data for this orbifold contains
two twisted sectors, both of which come with a multiplicity $5$
\cite{dgm}. We have shown the details of the graphical analysis
for this orbifold in figs. (\ref{fig9}) and (\ref{fig10}).
%%%%%%%%%%%%%%%%%%%%%%%%%%%%
%%%%%%%%%%%%%%%%%%%%%%%%%%%%%

We are now in a position to address the issue of the partition function
of perfect matchings for supersymmetric orbifolds of $\BC^3$. Consider
the orbifold $\BC^3/\BZ_3$. The partition function
\begin{eqnarray}
P &=& \prod_{i=0}^2 \left(x^{\frac{1}{3}} + \omega^i y^{\frac{1}{3}}
+ \omega^{2i}z^{\frac{1}{3}} \right)\nonumber \\
&=& x + y + z - 3x^{\frac{1}{3}}y^{\frac{1}{3}}z^{\frac{1}{3}}
\end{eqnarray}
with $\omega = e^{\frac{2\pi i}{3}}$
gives the twisted sector R-charge and the multiplicity of GLSM fields
for this orbifold. We interpret this result as the supersymmetric 
completion of the $\BC^2/\BZ_{3(1)}$ orbifold which is accomplished
in this case simply by adding an extra direction parametrized by $x$.
We find that this is a generic feature 
of the supersymmetric cyclic orbifolds of $\BC^3$, i.e the multiplicities
and twisted sector charges are obtained by making a supersymmetric 
completion of the corresponding non-supersymmetric $\BC^2$ orbifold. 

Next, let us consider the $\BC^3/\BZ_5$ orbifold. As is known, this model
has two blowup modes corresponding to the two marginal twisted sectors
of the closed string SCFT. We take the orbifolding action to be 
\footnote{Any other orbifold action is equivalent to this.}
\be
\left(Z_1, Z_2, Z_3\right) \to \left(\omega Z_1, \omega^2 Z_2, 
\omega^2 Z_3\right)
\ee  
where $\omega = e^{\frac{2\pi i}{5}}$. 
The partition function of perfect matchings is obtained by 
supersymmetrically completing the $\BC^2/\BZ_5$ orbifold. This can
be done in three ways, i.e either by completing the 
$\BC^2/\BZ_{5(1)}$, $\BC^2/\BZ_{5(2)}$ or the $\BC^2/\BZ_{5(3)}$ 
orbifold. The partition function 
\begin{equation}
P = \prod_{i=1}^2 \left(x^{\frac{1}{5}} + \omega^i y^{\frac{1}{5}}
+ \omega^{ki}z^{\frac{1}{5}} \right)
\end{equation}
with $k = 2,3,4$ (corresponding to the various non-supersymmetric
$\BC^2/\BZ_5$ orbifolds) gives the twisted sector charges and the 
multiplicities of the supersymmetric $\BC^3$ orbifolds. 
%%%%%%%%%%%%%%%%%%%%%%%%%%%%
%%%%%%%%%%%%%%%%%%%%%%%%%%%%%

A more interesting case is the supersymmetric $\BC^3/\BZ_7$ orbifold
that has two distinct actions
\begin{eqnarray}
\left(Z_1, Z_2, Z_3\right) \to \left(\omega Z_1, \omega^3 Z_2,
\omega^3 Z_3\right) \nonumber\\
\left(Z_1, Z_2, Z_3\right) \to \left(\omega Z_1, \omega^2 Z_2,
\omega^4 Z_3\right) 
\label{c3z7}
\end{eqnarray}
where $\omega = e^{\frac{2\pi i}{7}}$. In order to obtain the partition
function of perfect matchings, we could supersymmetrically complete the
non-supersymmetric orbifold $\BC^2/\BZ_{7(3)}$, in which case we will 
obtain the supersymmetric orbifold $\BC^3/\BZ_7$ with the orbifolding
action being the first one of eq. (\ref{c3z7}), or we could do the same
with the orbifold $\BC^2/\BZ_{7(2)}$ for which the second action of
eq. (\ref{c3z7}) is obtained (other orbifold actions of $\BC^2/\BZ_7$
give permutations of these results). 

Consider eq. (\ref{partition4}) with $n=7$ and $k=2$. This gives the 
orbifold $\BC^2/\BZ_{7(3)}$. Completing this partition function amounts
to the new function
\be
P = \prod_{i=0}^4\left(x^{\frac{1}{7}} + \omega^i y^{\frac{1}{7}}
+ \omega^{2i}z^{\frac{1}{7}}\right)
\ee
Expanding this, we obtain
\be
P = x + y + z - 7x^{\frac{1}{7}}y^{\frac{5}{7}}z^{\frac{1}{7}}+
14x^{\frac{2}{7}}y^{\frac{3}{7}}z^{\frac{2}{7}}+
- 7x^{\frac{3}{7}}y^{\frac{1}{7}}z^{\frac{3}{7}}
\ee
This gives the multiplicities and twisted sector charges of the
orbifold as in the first of eq. (\ref{c3z7}). Similarly, completing
eq. (\ref{partition4}) with $n=7$ and $k=3$ can be seen to reproduce
the twisted sector charges and the multiplicities of the second
orbifold action of eq. (\ref{c3z7}). A similar analysis can be
done for higher rank orbifolds of $\BC^3$ and the results follow
the same pattern as presented here. We now go over to a discussion
of supersymmetric orbifolds of the form $\BC^3/\BZ_n\times\BZ_m$.  

\subsection{Non-cyclic orbifolds of $\BC^3$}

\begin{figure}[h]
\centering
\epsfxsize=6.5in
\hspace*{0in}\vspace*{.2in}
\epsffile{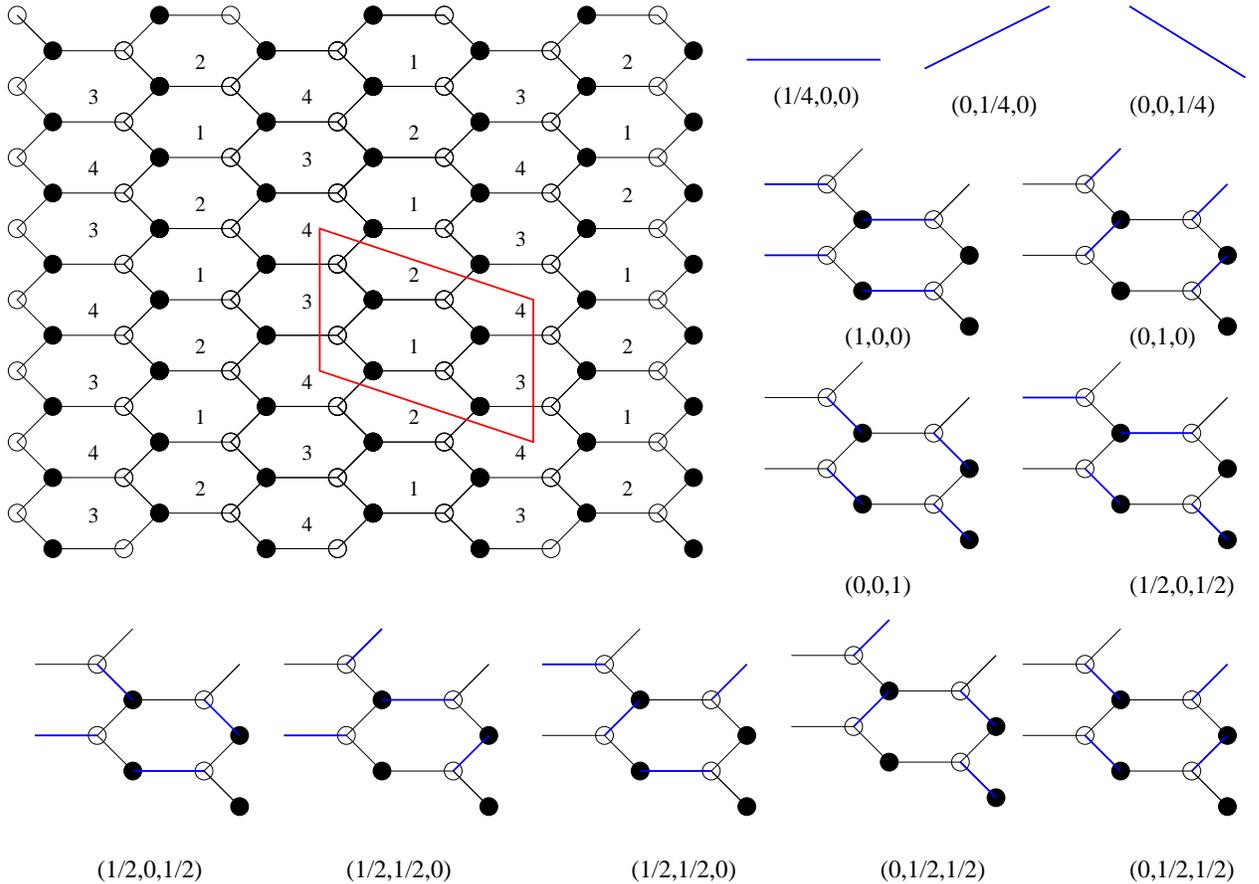}
\caption{\small The dimer model and perfect matchings for the 
supersymmetric orbifold $\BC^3/\BZ_2\times\BZ_2$.}
\label{fig8}
\end{figure}

In this subsection, we briefly discuss how our methods apply to the
orbifolds of $\BC^3$ of the form $\BC^3/\BZ_n\times\BZ_m$. We concentrate
on the simplest example, i.e for $n=m=2$, since the higher rank
orbifolds of this class have large multiplicities and are 
difficult to present graphically. For the  $\BC^3/\BZ_2\times\BZ_2$ orbifold,
we have presented the dimer model and the perfect matchings in
fig. (\ref{fig8}). Clearly, we see that our results of the last 
section can be adapted to this case with ease. In this case, we
weigh the edges by weights $\left(\frac{1}{2},0,0\right), 
\left(0,\frac{1}{2},0\right)$ and $\left(0,0,\frac{1}{2}\right)$.
This correctly gives us the twisted sector charges and the 
multiplicities of the orbifold \cite{fhh},\cite{ts3}. 

We can address the issue of partial resolution of singularities here.
This would corresponding to removing certain corners of the toric diagram.
This can be visualised in the following way : we choose the corner to
be resolved and identify the corresponding perfect matching. Then 
we choose any one of the edges of this matching (that has to be 
removed) and project out any other diagram that involves the said edge. 
%%%%%%%%%%%%%%%%%%%%%%%%%%%%
%%%%%%%%%%%%%%%%%%%%%%%%%%%%%
Eg. in fig. (\ref{fig8}), suppose we twisted sector corresponding
to the weight $\left(0,1,0\right)$. From the perfect matching corresponding
to this weight, we choose any edge and remove all diagrams that contains
this edge in its perfect matching. The twisted sector corresponding 
to the charges $\left(\frac{1}{2},0,\frac{1}{2}\right)$ is unaffected
whereas one each of the perfect matchings corresponding to the
charges $\left(\frac{1}{2},\frac{1}{2},0\right)$ and
$\left(0,\frac{1}{2},\frac{1}{2}\right)$ are seen to be ``resolved,''
and finally we get precisely the toric diagram with the multiplicities
of the SPP singularity \cite{fhh}. 
We leave a detailed study of the phenomena of partial resolutions
to a future work. 

\section{Conclusions}

In this paper, we have explored the connection between dimer models 
that naturally arise in the study of D-brane
gauge theories, and closed string theories. We have seen that
by weighing the edges of the perfect matchings of bipartite graphs, we
are able to give an interpretation of these perfect matchings as 
twisted sector states of closed strings on orbifolds. Since each 
perfect matching is a twisted sector state, this gives an explanation
of the dimer-GLSM correspondence, complementary to the methods used
in \cite{francovegh}. Further, we emphasize that this extends the
correspondence to non-supersymmetric orbifolds (of $\BC^2$) as well.
Note that in our description, we have not used the height function
as in previous works, but have established a direct connection with
twisted sector R charges via a counting of weighted edges. 

There are various issues yet to understand. The first question that one
might ask is whether there is an analogue of brane tiling corresponding to 
non-supersymmetric graphs. This issue can be addressed as follows.
Non-supersymmetric orbifolds of $\BC^2$ can be obtained from supersymmetric
orbifolds of $\BC^3$ by a method inverse to that of supersymmetric
completion \cite{crawreid}. This can be simply understood with an example.
Consider, eg. the supersymmetric orbifold $\BC^3/\BZ_5$, with the
orbifolding action being
\be
\left(Z_1, Z_2, Z_3\right) \to \left(\omega Z_1, \omega Z_2,
\omega^3 Z_3\right)
\ee
where $\omega= e^{\frac{2\pi i}{5}}$. A resolution of this 
supersymmetric $\BC^3$ orbifold can be obtained by considering 
three non-supersymmetric $\BC^2$ orbifolds, namely
$\BC^2/\BZ_{5(1)}$, $\BC^2/\BZ_{5(2)}$ and $\BC^2/\BZ_{5(3)}$. The toric
data of the resolution of $\BC^3/\BZ_5$ can then be constructed
by gluing the data for these three $\BC^2$ orbifolds. This procedure
has been dubbed the ``champions meet'' in \cite{crawreid}. Proceeding
in the reverse direction, it seems that an analogue of the brane tiling 
model for a non-supersymmetric $\BC^2$ orbifold can be obtained from
a supersymmetric $\BC^3$ brane tiling by removing one of the $\BC^3$
directions, which corresponds to removing one row from the charge 
matrix for the quiver diagram of the orbifold. However, as one can
check, this procedure does not provide a consistent tiling with
bipartite graphs, but rather one has to consider a tripartite graph.
A possible clue is that the gluing rules of \cite{crawreid}, when
translated in the language of graphs, might give us further hints as
to what is the analogue of brane tilings for non-supersymmetric 
orbifolds. This issue is under investigation. 

The issue of non-supersymmetric $\BC^3$ orbifolds (which might have 
terminal singularities) is also a interesting direction of future work. 
This will presumably involve construction of supersymmetric $\BC^4$ orbifolds,
and then using methods developed in this papr. 
We leave these issues to a future work.
  
\vspace{1cm} 
\noindent
{\bf Acknowledgments}\\

\noindent
It is a pleasure to thank Ami Hanany and for various 
useful discussions and email correspondence. Thanks are due to 
V. Subrahmanyam for lively discussions on dimer models.


\newpage
\begin{thebibliography}{99}
\bibitem{kast}
P. Kasteleyn, ``Graph theory and crystal physics,'' in {\tt
Graph theory and theoretical physics}, pp 43 - 110, Academic Press,
London, 1967.
\bibitem{kenyon}
R. Kenyon, ``An introduction to the dimer model,'' {\tt
math.CO/0310236}
\bibitem{han1}
A. Hanany, K. D. Kennaway, ``Dimer models and toric diagrams,'' 
{\tt hep-th/0503149}
\bibitem{han2}
S. Franco, A. Hanany, K. D. Kennaway, D. Vegh, B. Wecht, 
`` Brane dimers and quiver gauge theories,'' JHEP {\bf 0601} 2006, 096
\bibitem{han3}
S. Franco, A. Hanany, D. Martelli, J. Sparks,  D. Vegh, B. Wecht,
`` Gauge theories from toric geometry and brane tilings,''
JHEP {\bf 0601} (2006) 128
\bibitem{vafa1}
B. Feng, Y-H He, K. D. Kennaway, C. Vafa, `` Dimer models from mirror 
symmetry and quivering amoebae,''
{\tt hep-th/0511287}
\bibitem{wittenphases}
E. Witten, ``Phases of $N=2$ theories in two dimensions,''
Nucl. Phys. {\bf B403} (1993) 159, {\tt hep-th/9301042}
\bibitem{francovegh}
S. Franco, D. Vegh, `` Moduli spaces of gauge theories from dimer models: 
Proof of the correspondence,'' JHEP {\bf 0611} (2006) 054,
{\tt hep-th/0601063}
\bibitem{dm}
M. R. Douglas, G. Moore, `` D-branes, quivers, and ALE instantons,''
{\tt hep-th/9603167}
\bibitem{jm}
C. V. Johnson, R. C. Myers, ``Aspects of type IIB theory on ALE spaces,''
Phys. Rev. {\bf D55} (1997) 6382, {\tt hep-th/9610140}
\bibitem{he}
Y-H He, `` Lectures on D-branes, gauge theories and Calabi-Yau singularities,''
{\tt hep-th/0408142}
\bibitem{dgm}
M. R. Douglas, B. R. Greene, D. R. Morrison, 
``Orbifold resolution by D-branes,'' Nucl. Phys. {\bf B506} (1997) 84,
{\tt hep-th/9704151}
\bibitem{ts1}
T. Sarkar, ``Brane probes, toric geometry, and closed string tachyons,''
Nucl. Phys. {\bf B648} 497 (2003), {\tt hep-th/0206109}. 
\bibitem{fhh}
B. Feng, A. Hanany, Y-H He, 
`` D-brane gauge theories from toric singularities and toric duality,''
Nucl. Phys. {\bf B595} (2001) 165, {\tt hep-th/0003085}
\bibitem{reid}
M. Reid, ``Young person's guide to Canonical Singularities,''
Proc. Symp. Pure Math., {\bf 46} (1987) 345.
\bibitem{ag}
P. S. Aspinwall, B. R. Greene, `` On the geometric interpretation of $N=2$
superconformal theories,'' Nucl.Phys. {\bf B 437} (1995) 205,
{\tt hep-th/9409110}.
\bibitem{ts2}
T. Sarkar, `` On Tachyons in Generic Orbifolds of $\BC^r$ and Gauged Linear 
Sigma Models,'' JHEP {\bf 0702} (2007) 025, {\tt hep-th/0612046}.
\bibitem{han4}
B. Feng, S. Franco, A. Hanany, Y-H He, ``Symmetries of toric duality,''
JHEP {\bf 0212} (2002) 076, {\tt hep-th/0205144}.  
\bibitem{ts3}
T. Sarkar,  D-brane gauge theories from toric singularities of the form 
$\BC^3/\Gamma$ and $\BC^4/\Gamma$,'' Nucl. Phys. {\bf B595} (2001) 201, 
{\tt hep-th/0005166}.
\bibitem{crawreid}
A. Craw, M. Reid, ``How to calculate A-Hilb $\BC^3$,''
{\tt math.AG/9909085}.



\end{thebibliography}
\end{document}